**Title**

Quantum dot technology for quantum repeaters: from entangled photon generation towards the integration with quantum memories

Author(s)

*Julia Neuwirth\*, Francesco Basso Basset, Michele Beniamino Rota, Emanuele Roccia, Christian Schimpf., Klaus D. Jöns, Armando Rastelli, and Rinaldo Trotta\**

Affiliation(s)

Julia Neuwirth, Francesco Basso Basset, Michele Beniamino Rota, Emanuele Roccia, Rinaldo Trotta
Department of Physics, Sapienza University of Rome, 00185 Rome, Italy
E-mail: rinaldo.trotta@uniroma1.it, julia.neuwirth@uniroma1.it

Christian Schimpf, Armando Rastelli,
Semiconductor and Solid State Physics, Johannes Kepler University, 4040 Linz, Austria

Klaus D. Jöns
Department of Physics Paderborn University, 33098 Paderborn, Germany

Keywords: quantum repeaters, semiconductor quantum dots, quantum optics

Abstract

The realization of a functional quantum repeater is one of the major research goals in long-distance quantum communication. Among the different approaches that are being followed, the one relying on quantum memories interfaced with deterministic quantum emitters is considered as one of the most promising solutions. In this work, we focus on memory-based quantum-repeater schemes that rely on semiconductor quantum dots for the generation of polarization entangled photons. Going through the most relevant figures of merit related to efficiency of the photon source, we select significant developments in fabrication, processing and tuning techniques aimed at combining high degree of entanglement with on-demand pair generation, with a special focus on the progress achieved in the representative case of the GaAs system. We proceed to offer a perspective on integration with quantum memories, both highlighting preliminary works on natural-artificial atomic interfaces and commenting a wide choice of currently available and potentially viable memory solutions in terms of wavelength, bandwidth and noise-requirements. To complete the overview, we also present recent implementations of entanglement-based quantum communication protocols with quantum dots and highlight the next challenges ahead for the implementation of practical quantum networks.

**1. Introduction**

Modern information exchange based on optical communication is fundamental in present society. The current network structure, however, is inherently vulnerable to unwanted attacks, a hurdle that causes major privacy concerns as well as huge financial losses every year. A potential solution to this problem is provided by quantum networks in which information is encoded in single photons.[1,2] This approach promises stronger security by protecting data against eavesdropping—for example by offering future-proof data security in public key distribution[3,4]—exploiting the already vastly developed communication systems based on optical fibers and satellites. Once established, a network able to exchange quantum information could also provide novel functionalities such as enhanced node coordination for specific tasks or even support distributed quantum computation[2].

However, the bottleneck for long-distance communication is the scaling of the error probability with the length of the channel that connects transmitter and receiver. For example, in an optical fiber—the basis of the current optical communication systems—the probability for photon absorption and depolarization grows exponentially with the length of the fiber.[5] The detrimental effect can be even more severe for properties such as entanglement.[6] Therefore, an exponential number of trials for a successful transmission is needed. In classical communication, these problems are tackled with the use of repeaters along the channel, which amplify the signal and restore it to its original shape and intensity. Since signal amplification schemes cannot be applied to a quantum signal due to the no-cloning theorem,[7] alternative restoring methods are needed for quantum networks, and the idea to equip the nodes of a segmented transmission channel with quantum repeaters has been explored.

In a nutshell, a quantum repeater is a technology that enables the distribution of quantum entanglement among the distant (and uncorrelated) nodes of a quantum network without suffering from unbearable signal losses. It relies on entanglement resources, projective measurements and, arguably,[8–10] quantum memories. Different strategies are currently being followed for its implementation: One can use dissimilar quantum systems for the entanglement resource and the memory[11,12], the very same platform for both[13] (often using spin-photon interfaces in which the spin qubit is used as quantum memory and the photonic qubits to interface the different nodes[14–16]) or even memory free-schemes that rely on measurement-based operations on complex cluster states.[10,17–19] While all the different approaches have their own advantages and disadvantages, the construction of functional quantum repeaters will most likely require merging the different concepts being developed.[9] Here, to better introduce this work, we focus on hybrid schemes relying on near-ideal sources of entangled photon pairs[20–22] based on quantum dots and external quantum memories, and we start out discussing in more detail how such a quantum repeater[23] would work, see **Figure 1**.

Entangled photon pairs from quantum emitters are generated at intermediate nodes of a quantum network. The distribution of entanglement is then obtained using a quantum relay, i.e., by connecting two adjacent segments via a procedure known as entanglement swapping.[24] One photon from each entangled-photon pair is used to implement a projective measurement via photodetectors—the so-called Bell state measurement—and thus to finally project the remaining furthermost photons on an entangled state. While the cascaded sequence of swapping operations enables entanglement distribution among the different nodes, the probability that all photons propagate across the entire network is exactly the same of a photon propagating across a communication channel with the same overall length (despite a potential gain in signal-to-noise ratio[25]). Here, quantum memories come into play: When entanglement swapping is successfully performed on a block of the channel, the output photons are temporarily stored until the operation is also performed in the adjacent blocks. Thus, quantum memories are needed to store entanglement and to time its distribution in case entanglement swapping failed at some point of the network. It is important to emphasize that it is exactly the combination of quantum relay and memory that overcomes the difficulty associated with the exponential decay of the signal, allowing for, in principle, an overall communication fidelity very close to unity with a communication time growing only polynomially with transmission distance.[23] If non-ideal swapping operations decrease the overall entanglement, purification schemes can be added and applied to regain a high degree of entanglement.[26]

Despite the simple—yet ingenious—underlying concept, each step in the operation of a quantum repeater needs to be accurately performed for the whole technology to be advantageous, and this poses high demands on its basic hardware blocks. First, entanglement swapping requires photon pairs to be emitted with high efficiency, very high degree of entanglement and, due to the Bell state measurement, high photon indistinguishability.[27] Therefore, stringent requirements are set for the entangled photon sources. So far, sources based on parametric down conversion[28] have been used for the majority of photon-based entanglement experiments.[27,29,30] These sources provide entangled photon states with an outstanding degree of entanglement by taking advantage of nonlinearities in the used material.[28] However, a high degree of photon entanglement is achieved at the cost of limiting the brightness of the source. This is because there is a non-negligible probability to emit multiple photon pairs in a single excitation cycle, making it vulnerable to errors in the quantum relay operation[31] and hence lowering the overall efficiency of the quantum repeater scheme.

For these reasons, alternative entangled photon sources have been investigated in the last two decades. One promising entangled photon source, with the potential of generating entangled photons on demand and meeting the stringent requirements of the quantum repeater schemes are semiconductor quantum dots (QDs). QDs can generate polarization-entangled photon pairs via the biexciton-exciton cascade[32] with extremely low multi-photon emission probability,[33,34] high degree of entanglement,[35] photon-indistinguishability,[21,22] wavelength-tunability[36] as well as with nearly on-demand generation.[37] Additionally, QDs can be easily integrated within photonic chips and are compatible with the semiconductor technologies. These features hold great promise for the construction of solid-state-based quantum repeaters.

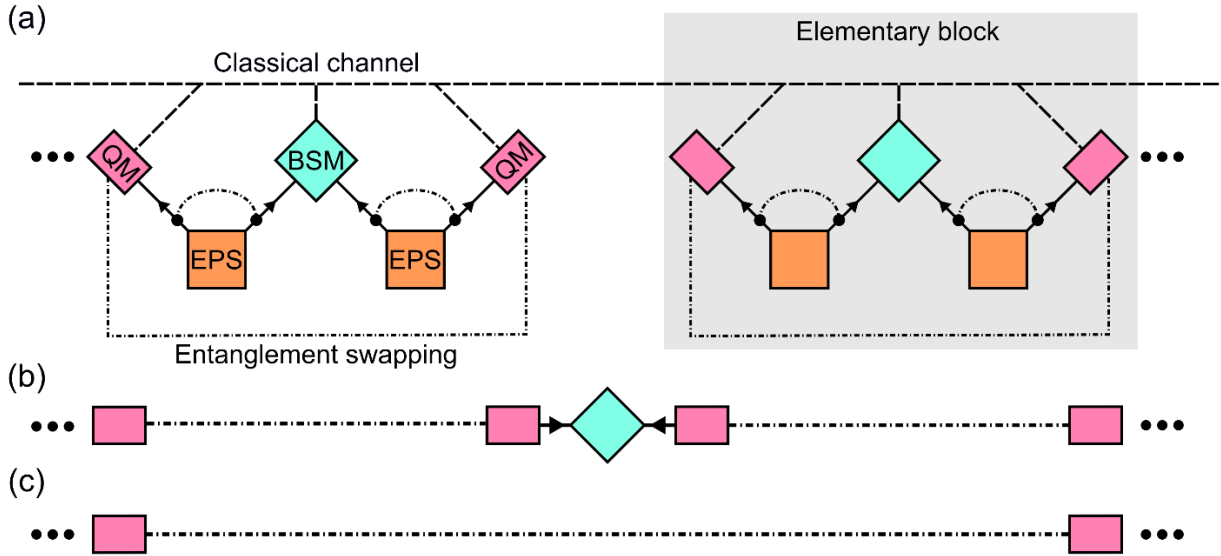

*Figure 1: Memory-based quantum repeater scheme. a) Two adjacent elementary blocks of a quantum repeater. On the right, we highlight the whole block with a grey rectangle, while on the left we label and emphasize its components. Two entangled-photon sources (EPS) are each interfaced with two quantum memories (QMs). Each QM is for storing one photon from an EPS. By performing a Bell state measurement (BSM) on the two inner photons in the diagram, the entanglement (represented by the dashed lines) is teleported to the outer photons, the ones stored in the QMs. b) Once the entanglement swapping operation is completed in two adjacent blocks, a BSM can be performed between the closest QMs, c) resulting in entanglement distribution to a further distance.*

The second set of requirements, which is essential to the viability of the presented quantum repeater protocol, concerns the storage and retrieval of photon states. The entanglement that is successfully shared among pairs of intermediate nodes needs to be stored for a sufficiently long time to allow for entanglement swapping steps to be completed in adjacent nodes, measurement outputs to be communicated via a supporting classical channel, and local operations to be performed on the entangled nodes. Quantum memories, able to comply with this demand while introducing a limited amount of errors in the process, are essential to ensure the polynomial time scaling of the quantum repeater protocol.[38] For this hardware element as well, several different technologies can be considered and the research to identify the best solution is still active and open. In a quantum memory, the photonic state can be transformed into stationary qubits, as it is difficult to store photons for a reasonably long time without absorption or degradation of the state. Earlier quantum memory proposals rely upon entanglement of single atoms obtained through single-photon interference at photodetectors.[39,40] Other approaches, in contrast, are based on collective excitations in atomic ensembles[41,42] or rare-earth crystals coupled to photonic crystal cavities[43] to exploit the enhanced coupling to light. High standards such as high efficiency of the strong coupling between atoms and photons,[44] long storage times,[11] and bandwidth matching with other components in the quantum network[45] are requested and constitute the figures of merit of a quantum memory performance.

In this work we focus on QD-based entangled photon sources and summarize the recent developments towards their application in real-life quantum repeaters. The first two sections discuss the topic of entanglement photon generation with QDs, with focus on the fine structure splitting, the excitation methods, as well as on the techniques that alleviate the remaining imperfections. Special attention is given to Al-droplet etched GaAs quantum dots. Next, a selection of innovative concepts to improve extraction efficiencies and photon indistinguishability up to the level in which QDs can meet the stringent requirements of quantum repeaters are discussed. In the following section, several promising quantum-memory technologies and protocols, suitable for interfacing with photons from QDs, are introduced, together with some preliminary studies in that direction. Finally, in the last section, we present seminal applications of QD-based photon sources to quantum communication protocols preliminary to the quantum repeater, namely quantum teleportation and entanglement swapping. Moving from our experimental demonstration and the research state-of-the-art, we give a final outlook on the future challenges to ultimately realize quantum repeaters based on entangled photon pairs from QDs.

## 2. Main

Semiconductor QDs are nanostructures consisting of several thousand atoms, self-assembled during epitaxial growth. Due to the confinement of the motion of charged carriers along all spatial directions, discrete energy levels result, similar to real atoms. The radiative cascade from a three-level system formed by the biexciton (two electron-hole pairs bound via Coulomb interaction), the exciton (one electron-hole pair) to the ground state (empty QD) can lead to the emission of two photons that are entangled in the polarization degree of freedom. In order to achieve a very high degree of entanglement, it is fundamental to have at hand QDs with high electronic symmetry. Thus, we begin our discussion explaining how to accomplish this task by controlling the exciton fine structure splitting.

### 2.1 Fine Structure Splitting

The capability to use the biexciton-exciton-ground state cascade to generate polarization entangled photons can be immediately recognized taking into account that (*i*) the exciton is built up from electrons and heavy-holes having spin quantum number of $S_{e,z}$ = ±1/2 and $J_{hh,z}$ = ±3/2;[46] (*ii*) the polarization of the emitted photons is directly connected to the total angular momentum projection *M* of the recombining electron-hole pair and only transitions with a momentum variation *ΔM* of ±1 are optically active; (*iii*) in QDs with high structural symmetry, e.g. $D_{2d}$, the bright excitons are doubly degenerate and characterized by *M* of ±1. The biexciton state is in a symmetric linear superposition of these angular momentum states,[47] with null total angular momentum projection. This implies that in ideal QDs the radiative cascade can take place via two decay paths, i.e., via the emission of a right- followed by a left-circularly polarized photon, or vice versa. If the two paths are indistinguishable, as illustrated in **Figure 2a** with the dashed intermediate state, the resulting two-photon state is entangled in polarization, precisely in the maximally entangled ɸ⁺ Bell state.

However, the real situation is quite different from this ideal scenario, in particular concerning requirement (iii). Despite remarkable improvements in fabrication techniques, inevitable fluctuations in composition, size, shape, arrangement in the host matrix, and intermixing with the substrate and the cap material make the fabrication of symmetric QDs a mere theoretical idea.[48,49] The deviation from an ideal QD with $D_{2d}$ symmetry to a real QD with $C_{2v}$ or even $C_1$ symmetry induces, via the anisotropic electron-hole exchange interaction, a fine structure splitting (FSS) and a coupling of the two bright excitonic states,[46,50] as shown in **Figure 2a** (specifically by the solid-line exciton levels). This has a dramatic impact on the resulting two-photon entangled states. When the FSS is larger than the radiative linewidth of the exciton transition, i.e. about 3 µeV for GaAs QDs,[35,51] there is an evolution of the entangled state over time[52] and the degree of entanglement, quantified by the time-averaged fidelity to a maximally entangled Bell state, is strongly reduced.[52]

Various filtering techniques can be used to minimize the effect introduced by FSS. For example, narrow spectral filtering of the biexciton and exciton photons can be applied to improve the fidelity of the entangled state.[53] Another related approach is based on temporal post-selection of the emitted photons,[54–56] which has demonstrated to achieve high entanglement fidelities.[57] However, it has to be emphasized that post-selection schemes inevitably induce severe photon-pair losses, hamper on-demand generation of entangled photons and, more in general, the real possibility of using QDs for applications. In contrast to post-selection schemes, external perturbations,[58,59] such as electric,[60–62] magnetic,[52] and strain[63–65] fields, provide the possibility to compensate for the FSS by directly modifying the QD electronic structure.

First approaches using single external perturbations demonstrate a minimization of the FSS to small, albeit non-zero values. This is due to the fact that the QD asymmetry has to match the applied perturbation,[66,67] and only in very special cases complete suppression of the FSS could be achieved. A universal cancellation of the FSS, however, can be accomplished by the simultaneous application of two external perturbations applied simultaneously. This is due to the fact that two perturbations are needed to achieve full control over the two QD parameters: the magnitude of the FSS and the polarization direction of the exciton emission.[68] Reversibly controlled and erased coupling between the bright excitons in arbitrary QDs has first been shown by Trotta *et al.* using strain and electric field.[69] However, the latter field inevitably affects the oscillator strengths of optical transitions, and alternative approaches are needed to reshape the QD electronic structure without affecting the QD parameters that are relevant for efficient entangled photon generation.

Based on this, two "tuning-knobs" concepts have been proposed by Wang *et al.*[70] and Trotta *et al.*[36], as well as Chen *et al.*[71] and could demonstrate that it is possible, not only to use strain to suppress the FSS without affecting the brightness of the source, but also to achieve control over the photon emission energy. The former proposed a device built on a piezoelectric lead zirconate titanate (PZT) ceramic stack with the [100], [010], [001] axes of the QD sample aligned with the polar (z,y,x) axis of the PZT. Applying two independent voltages introduces two independent in-plain strain components that are used to cancel the FSS of an arbitrary QD. Additionally, stress can be applied from the top to bottom of the sample to tune the emission energy. This arrangement, however, requires the technologically challenging use of a transparent strain transmitter between the sample and the top stressor. The latter concept introduced by Trotta et al.[36] is built on a micromachined piezoelectric structure featuring six legs aligned at 60° with each other, as shown in **Figure 2b** . These legs can be independently contracted by applying a voltage. In order to cancel the FSS of an arbitrary QD, a nanomembrane containing the QDs is bonded onto the micromachined piezoelectric structure and by applying independent voltages to opposite-facing legs quasi-uniaxial stresses can be achieved. The excitonic degeneracy can be restored by using one leg to align the polarization axis of the exciton emission along the direction of a second leg, which is then used for the complete compensation of the difference in energy of the two bright exciton eigenstates, as shown in **Figure 2c**. Consequently, two legs of this three-leg design can be used for FSS cancellation, whereas the third one allows for additional tuning of the exciton emission energy. As the top face of the membrane is free, emitted photons can be easily collected by a microscope objective. This approach, which has been experimentally demonstrated for the first time in 2016[72], emphasizes the possibility of successfully compensating for structural anisotropies introduced by the fabrication techniques without the loss of emitted photons and, more importantly, without spoiling the degree of entanglement as well as on-demand emission.

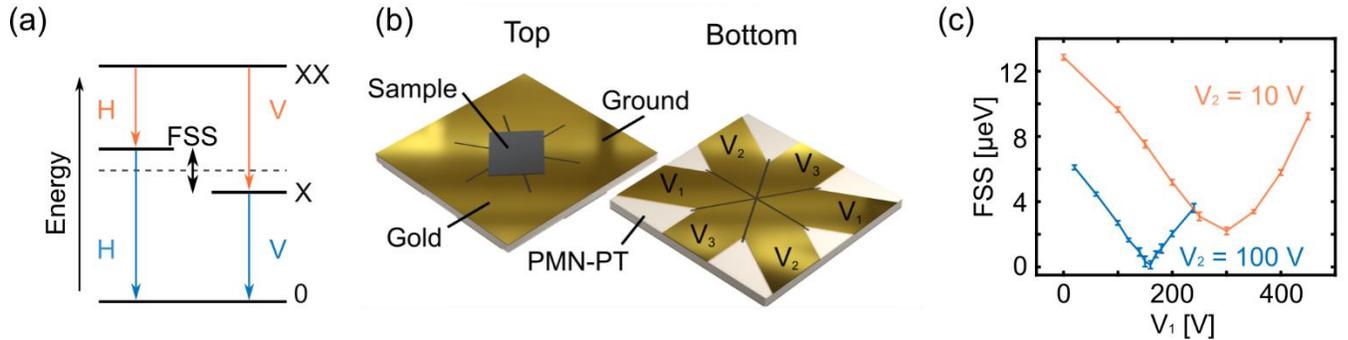

*Figure 2: Fine structure splitting and the possibility for its cancellation. a) Energy diagram of the biexciton-exciton radiative cascade. The electron-hole anisotropic exchange interaction results in an energy splitting in the intermediate exciton state, called fine structure splitting (FSS, solid-line X levels). When the FSS vanishes (dashed-line X levels), the QD excited in the biexciton state generates a polarization-entangled photon pair by decaying through the two equivalent optically active paths. b) Schematic illustration of the six-leg piezoelectric device. Top and bottom view of the structure. The piezoelectric substrate (PMN-PT) has three cuts, leaving six individual legs that are connected pairwise on the bottom (indicated by the three voltages $V_1$-$V_3$). The top surface is connected to the ground. The sample (gray layer) is connected to the top surface of the six-leg device. c) Minimization of the FSS as described in the text by applying $V_2 = 10$ V on leg 2 and tuning the voltage on leg 1. By adapting the applied voltage on leg 2 to $V_2 = 100$ V the FSS can be cancelled. Adapted from Ref.[20]*

## 2.2 Excitation methods

On-demand sources of entangled photon pairs are essential components in quantum communication protocols and semiconductor QDs are currently considered as one of the most promising sources for generating on-demand entangled photon pairs.[37] Since the radiative cascade from the biexciton state via the exciton state to the ground state is used for this purpose, a topic of particular relevance is the pumping of the system into the biexciton state.[32] This is the focus of the next section, in which we deal with scheme for electrical and optical pumping, non-resonant and two-photon resonant excitation.

Electrically driven sources of entangled photons are of special interest as the absence of an excitation laser would drastically reduce the complexity of quantum networks.[73] Also, pulsed mode excitation with controllable rate can be implemented in a rather straightforward manner, giving access to fast generation rates of entangled-photon pairs—a highly desirable feature for fast data-rate quantum information processing. In fact, this mode of operation has been demonstrated by embedding QDs in the intrinsic region of p-i-n light-emitting diodes.[74,75] The clock rate of QD-based entangled photon diodes has been pushed up to 1.15 GHz by Mueller et al. in 2020,[76] while achieving an entanglement fidelity of 0.795. These figures of merit are mainly dictated by the radiative lifetime of the emitter and by the impact of re-excitation processes. Electrical carrier injection has also been demonstrated to be compatible with the strain-tuning techniques for FSS control presented in the previous section, with the realization of a functional device operating at a pumping rate of 400 MHz.[77]

Despite the potential advantages, electrically driven sources currently suffer from a main drawback. With electrical injection, electron-hole pairs are generated in the semiconductor matrix around the QD and subsequently diffuse into its potential well. This relaxation process can result in the excitation of several different excitonic complexes, so that the creation of a biexciton state can be considered a probabilistic process. As previously mentioned, it can also lead to re-excitation, which decreases the measured entanglement fidelity of the source. Finally, the carrier relaxation to the lower confined energy level introduces time jitter, which, as we will discuss in a following subsection, has a negative impact on photon indistinguishability.

Optical excitation offers the possibility to avoid these hurdles, provided that a resonant excitation scheme is used. When the laser energy is adjusted such that carriers are excited in the barrier material or in the wetting layer of the QD, spectral features similar to the case of electrical injection are observed. The photoluminescence spectrum of a single GaAs QD under non-resonant excitation is illustrated in **Figure 3a**. The exciton line on the high-energy side of the spectrum is very pronounced, but the biexciton line can hardly be singled out from the many possible charged states that emit close to the biexciton energy.

A way to resonantly excite the biexciton state under compliance with the electric-dipole selection rules consists in using a two-photon absorption process.[37,78–80] The laser energy is set between the exciton and biexciton emission energy and, in QDs with a sufficiently large and positive biexciton binding energy, it cannot directly populate the exciton state with experimentally achieved repetition rates up to 1 GHz[81]. Additionally, the probability of populating other states or charge configurations is also drastically reduced. As an example, the photoluminescence spectrum of the same GaAs QD from **Figure 3a** is shown under two-photon resonant excitation in **Figure 3b**. Rabi oscillations of exciton and biexciton emission intensities, such as those reported in **Figure 3c**, demonstrate the coherently-driven operation, and near-unity population probability of the biexciton state, 0.98(7),[37] after π-pulse can be achieved. Two-photon excitation also minimizes recapture processes, as supported by the experimentally observed strong antibunching.[34] **Figure 3d** reports the second-order autocorrelation function of the biexciton line from Schweickert et al.[34] yielding a record-low value of normalized coincidences at zero-time delay of 7.5(1.6) $10^{-5}$. The suppression of recapture processes also allows to achieve high entanglement fidelities, with demonstrated values up to 0.978(5).[20]

To complete this section, it is important to mention that there exists another effective method to pump the biexciton state: phonon-assisted two-photon excitation. In this scheme the excitation laser is detuned to higher energy, with respect to the two-photon resonant excitation energy, to address the vibrational modes coupled to the biexciton state. The exact detuning for optimal state preparation is determined by the QD structure and the surrounding material deformation potential.[82] In contrast to the strict two-photon excitation, this scheme provides a stable population probability for fluctuating excitation laser power. However, if one aims at simultaneously optimizing the degree of entanglement, photon indistinguishability, single-photon purity, and on-demand generation, two-photon resonant excitation has so far proven to be the best solution and is thus often used for entangled photon generation in QDs.

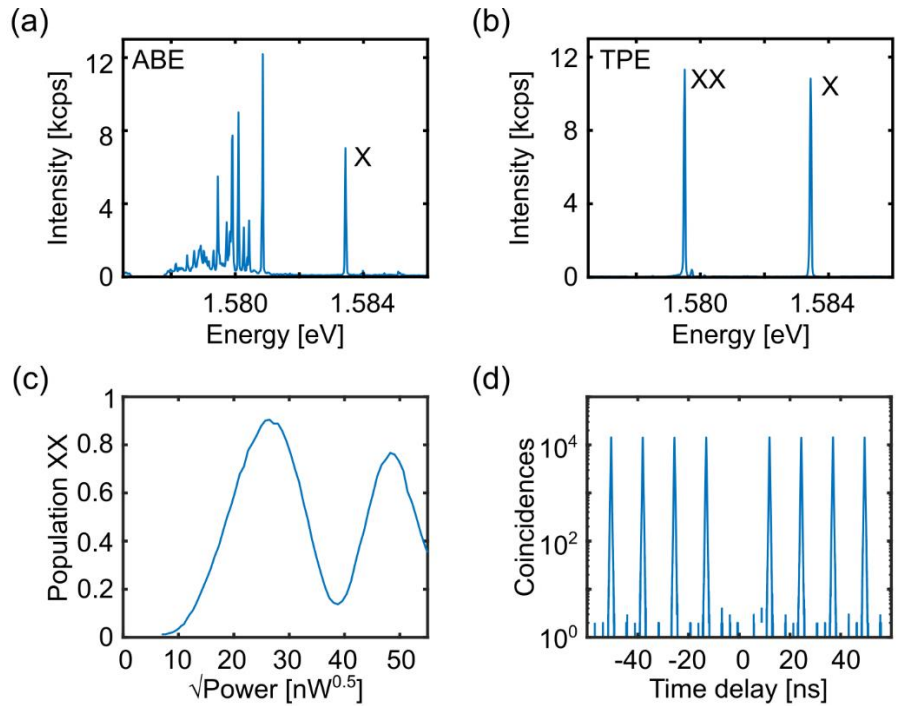

*Figure 3: Typical QD excitation schemes for populating the biexciton state. a) Micro-photoluminescence spectrum of a GaAs QD under non-resonant above-band excitation (ABE). The neutral exciton emission line is labelled as X. The biexciton line (XX) lays within the bundle of emission lines at lower energy and, thus, cannot be distinguished. b) Spectrum of the same QD as in (a) under resonant two-photon excitation (TPE). The laser is set to a virtual level that is half the XX energy and not resonant with the X energy because of the different coulomb binding of the X and XX levels. The exciton and biexciton lines are clearly visible and are the dominant emission components. c) Rabi oscillation of the XX emission intensity vs. laser power. The emission intensity is normalized and multiplied by the estimated preparation fidelity of that QD. d) Second-order autocorrelation function measured for the XX emission. Adapted from Ref. [34]*

## 2.3 Brightness

Another relevant factor for the successful implementation of an entangled photon source in quantum repeaters is its brightness. A common definition of the brightness of the source, among the wide range,[83] is the probability of collecting a single photon with the first lens of the collection optics upon an excitation pulse. Semiconductor QDs can in principle generate deterministic photon pairs close to unity efficiency without negatively influencing the entanglement fidelity.[37,80] The main limiting effect for brightness in semiconductor QDs is a small extraction efficiency, related to the fact that QDs are embedded in a host matrix, like GaAs or AlGaAs, which typically has a high refractive index $n$, e.g. 3.5 in GaAs. Therefore, the extraction efficiency is limited for planar as-grown samples due to total internal reflection, which allows only a small fraction ($1/(4n^2)$) of the emitted light to have the chance of exiting the sample.[84] Taking into account the finite numerical aperture (NA) of the collection optics and interface reflection, generally, only less than 1 % of the emitted light can be collected from the top surfaces of planar unprocessed samples.[85]

Over the years, several methods have been employed to improve the brightness of QD-based sources. One method is based on solid immersion lenses. With the use of a Weierstrass sphere with $n$ = 1.88 and a collection system NA of about 0.53 a theoretical extraction efficiency of up to 11% can be achieved,[85] however with the unrealistic assumption of perfect transmission at all interfaces. For a semi-spherical solid immersion lens with $n$ = 3.5 and a collection system NA of about 0.42, an extraction efficiency of 65(4)% was recently demonstrated.[86]

A second approach based on photonic structures embeds the QDs in a planar lambda-cavity either defined by Distributed Bragg gratings (DBRs), metal mirrors or a hybrid DBR/metal mirror system. The mirrors reflect the produced field from the emitter back to the emission site. Thus, the reflected field drives the emitter and enhances the emission of the source, in case reflected and emitted field are in phase.[85] **Figure 4a** illustrates an example of the combination of a glass-made solid immersion lens with a DBR structure to enhance the extraction efficiency. However, the single-photon extraction efficiency, with measured values of approximately 12%, is still relatively modest.[87]

A decisive step towards the development of bright QD single-photon sources was achieved by exploiting cavity quantum electrodynamics. By establishing an additional lateral optical confinement via etching micropillars out of the planar cavity, single-photon extraction efficiencies as high as 79% have been demonstrated,[88] whereby it has to be noted that additional spectral fine-tuning is performed using temperature. Very recent results have shown that an optical microcavity can also be realized without the microfabrication of lateral structures, using an open cavity design.[89] This has led to an extraction efficiency of 83%, together with an excellent single-mode coupling efficiency which results in an overall probability of having a single photon at the fiber output of 57%.

These microstructures, even though presenting exceptional extraction efficiencies for single photons, are less suitable for entangled photon pair generation as the energy difference between the biexciton and exciton typically exceeds the resonance width of these cavities. A possible solution to this problem is a photonic molecule design formed by two fused micropillars with a controlled inter-center distance coupled to a single QD.[90] This approach successfully demonstrated a pair extraction efficiency of 12%, but requiring complex microfabrication. A potentially simpler alternative relies on the use of an optical waveguide for collecting the light in an efficient way without the need of a microcavity. This satisfies the required broadband operation and has been proved compatible with the generation of entangled photons from a QD embedded in a nanowire.[91] The reported pair extraction efficiency, while in principle being improvable to near-unity values, is currently limited to 3%. Moreover, a waveguide alone would not exhibit potential benefits deriving from the Purcell effect.

An in-between more sensible approach to enhance the extraction efficiency of entangled photon sources makes use of cavities with a broadband extraction efficiency and moderate Purcell factor. In this context, circular Bragg gratings, also called bullseye cavities,[92] are particularly promising. An example image of a circular Bragg grating is illustrated in **Figure 4b**. The microstructure is accurately patterned around selected single QDs. As a result, photons emitted by the QD in the center of the circular Bragg grating will be directed perpendicular to the concentric trenches. When combining these bullseye cavities with highly efficient broadband reflectors, outstanding pair collection efficiencies can be achieved. Under a pulsed two-photon resonant excitation scheme single-photon extraction efficiencies of 85(3)% and pair extraction efficiencies of up to 65(4)% were obtained in GaAs QDs,[21] and similar results were obtained with InGaAs QDs[22]. Theoretical estimations predict collection efficiencies above 90% in a bandwidth of around 33 nm, and, additionally, Purcell factors above 5 within a bandwidth of around 8 nm (20 at the center wavelength).[21] The developed design and the possibility of using different substrates furthermore allow the combination with piezoelectric actuators[21,93] for the cancellation of the FSS, as well as for wavelength tuning. This combination would not only give access to superior extraction efficiency and subsequently bright single and entangled photon sources based on semiconductor QDs, but would also ensure high entanglement fidelity and indistinguishability—necessary properties for most quantum information experiments based on entangled photon sources.

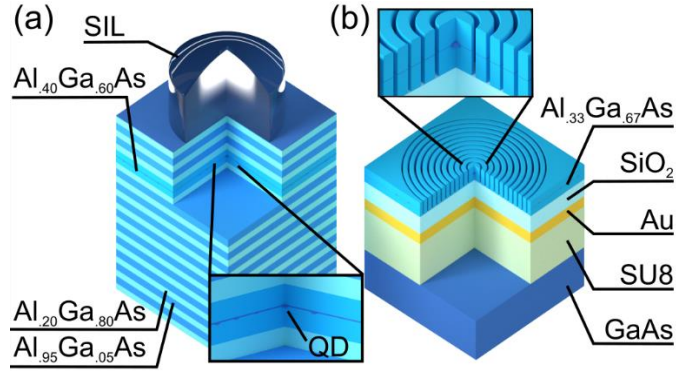

Figure 4: Sample structure to improve the extraction efficiency. a) Sample structure with DBR (blue and light blue alternating layers) and solid immersion lens (SIL). Close up of the QD in the inset. b) Sample structure with a bottom broadband reflector, made by a gold (Au) plus an oxide (SiO$_2$) layer, and a second-order circular Bragg grating. Close up of the QD in the inset.

## 2.4 Indistinguishability

Numerous applications in photonic quantum technologies rely on quantum interference of deterministically generated single or entangled photons. Such schemes include Bell state measurements,[94] quantum teleportation,[30] post-selective production of polarization-entangled photons,[95] linear-optics quantum computation,[96] and boson-sampling[97]. As mentioned above, light sources for quantum repeaters should emit highly indistinguishable and entangled photons. Two photons are indistinguishable if they are identical in their spatial, spectral, polarization, and temporal modes. The photon indistinguishability can be verified by measuring the two-photon interference visibility in a Hong-Ou-Mandel (HOM) experiment.[98] An example of a HOM experiment for co-polarized exciton photons coming from subsequent biexciton-exciton radiative cascades from the same GaAs QD under resonant excitation, is shown in **Figure 5a**. The coincidence counts at zero-time delay indicate a non-perfect indistinguishability between the two interfering photons. This is caused by many factors,[99] namely time jittering,[100] phonon-induced dephasing,[101] fluctuating magnetic and electric fields.[102,103]

Decoherence introduced by time jittering is caused by non-resonant or quasi-resonant excitation schemes,[100,104] because the stochastic relaxation process from a higher excited state to the lowest-energy level via nonradiative processes leads to an uncertainty in the emission time, which degrades the indistinguishability. Therefore, time jittering can be largely overcome by using resonant excitation.[105] Phonon-induced dephasing, on the other hand, results from the inevitable coupling of the QDs to the vibrational modes of their host lattice. Inelastic exciton-phonon scattering produces detuned, distinguishable photons in sidebands by exchanging energy with the QD.[106] Also, elastic phonon-exciton scattering through virtual excitations introduces a broadening of the zero-phonon line, which decreases the photon indistinguishability.[107] The latter can be mitigated by operating the source at lower temperatures than the characteristic phonon energies. Additionally, phonon-induced effects can be significantly reduced by embedding the QD in a photonic cavity that selectively enhances zero-phonon emission processes through the Purcell effect.[108]

Charge and spin noise, the last predominant indistinguishability degradation mechanism in QDs, arises from random occupations of the available electronic states in the surroundings of the confined excitons and leads to fluctuations in the local electric field. This can induce shifts in the optical transition energy of a nearby QD through the quantum confined Stark effect and can induce spin dephasing through the spin-orbit interaction.[102] Therefore, the linewidth of a self-assembled QDs can be significantly increased above the natural linewidth limit, thus deteriorating the achievable indistinguishability. It should be mentioned that slow fluctuations in the local magnetic field or fluctuations in nearby crystal impurities do not necessarily affect the indistinguishability of photons emitted subsequently from the same source.[109,110]

Schöll et al.[111], e.g., recently demonstrated that, under cross-polarized pulsed resonance fluorescence, quantum dots fabricated using droplet-etching epitaxy—previously introduced as state-of-the-art entangled photon emitters—do not suffer from dephasing mechanisms at short time scales. Using this excitation technique, a value of $V_{raw}$ = 0.95(5) for an on-demand single-photon source was achieved, without the need for any enhancement technique such as microcavities. However, this excitation technique is incompatible with entangled photon pair generation, as two-photon resonant excitation is needed to coherently populate the biexciton state. In this case, due to the lifetime ratio between the biexciton and exciton states, is the raw HOM visibility usually limited to about 70%,[112] even if it can be improved using the Purcell effect. Improved photon indistinguishability was indeed recently demonstrated using circular Bragg reflectors.[21,22] In particular, Liu et al.[21] reported values of indistinguishability of 0.901(3) and 0.903(3), for exciton and biexciton photons respectively.

These high indistinguishability values were all achieved by interfering two photons from the same QD. For quantum repeater protocols, however, the interference of two photons from remote QDs is paramount. The matter has seen limited investigation on entangled photon sources, although the considerations from several studies on single-photon sources[113–118] are largely transferable. Using two-photon resonant excitation, the best remote visibility for cascaded photons to date has been achieved by Reindl *et al.* in 2017,[119] who used GaAs QDs embedded in a planar DBR cavity under phonon-assisted two-photon excitation. **Figure 5b** shows the two-photon interference of co- (blue) and cross- (orange) polarized photons from two remote GaAs QDs under phonon-assisted excitation. From the experimental data, a visibility of V = 0.51(5) has been extracted. Despite this achievement, higher visibilities are needed to successfully implement quantum repeater schemes (see Section 2.6). However, this is an enormous challenge. First, single QDs with radiative linewidth as close as possible to the Fourier limit need to be fabricated. So far, this has been demonstrated only by a few groups who also reported high HOM visibilities with photons from the same QD but pumped with pulses separated by microseconds.[89,120,121] Yet, high visibility between separate narrow-linewidth QDs has only been observed in the coherent scattering regime of operation,[122] which is incompatible with the nearly deterministic generation processes we are considering here. Second, every QD exhibits its specific characteristics—in terms of emission energy, temporal shape of the photon wavepacket, as well as a different charge-fluctuating environment—degrading photon indistinguishability. While external perturbations[72] can be used to mitigate some of these deleterious effects, so far no experimental group has attempted quantum teleportation experiments with photons from remote QDs. This will likely require spectral filtering[123] and temporal post-selection of the emitted photons[124].

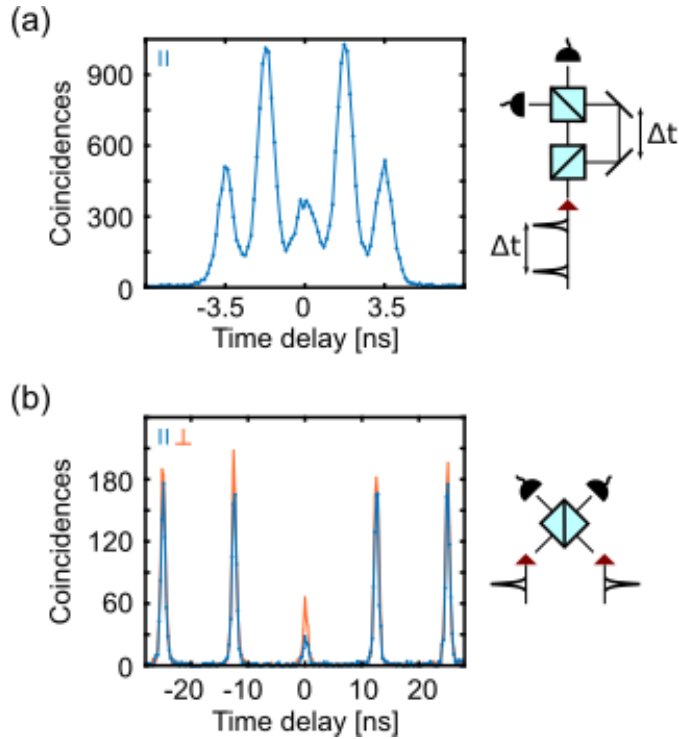

Figure 5: a) Two-photon interference coincidence histogram from co-polarized photons emitted by the same GaAs QD with a delay of 1.8 ns, under resonant two-photon excitation. Adapted from Ref.[119] b) Two-photon interference coincidence histogram with co-polarized (blue) and cross-polarized (orange) photons emitted by different GaAs QDs, strain-tuned to the same emission energy and excited by phonon-assisted two-photon excitation. Adapted from Ref.[119].

## 2.5. Quantum memories

When we have chosen an entangled photon source, the next building block for the quantum repeater is the quantum memory. The various quantum memory approaches for quantum repeater schemes are quite diverse, including solid-state atomic ensembles in rare-earth doped crystals,[125] NV centers in diamond,[126] semiconductor quantum dots,[127] single trapped atoms[128] and, room-temperature[129] and cold atomic gases[130]. The efficiency, the fidelity, and the storage time are the figure of merit with the largest impact on the quantum repeater performance. In addition, of particular importance for hybrid systems is the bandwidth matching between photon source and memory. Often, highly efficient memories exhibit a narrow bandwidth which is incompatible with QDs. The efficiency of a quantum memory is quantified by the probability to re-emit a photon after storage. High recall efficiency is favorable because it increases the success rate of entanglement distribution and thus makes scaling of quantum networks easier. In atomic ensembles, the efficiency can in principle be close to one due to collective interference effects.[44] For single atomic systems, on the other hand, the efficiency of recovering a single emitted photon is typically small but can be enhanced by optical cavities.[131] Another promising approach are rare-earth doped crystals. By embedding them in perfectly impedance-matched cavities unity efficiency should theoretically be achievable.[132] The second crucial performance characteristic is the fidelity, which measures the overlap of the quantum state of the retrieved photon after storage with the input state. For practical quantum memories the fidelity needs to be higher than the no-cloning limit to render undetected eavesdropping impossible.[133,134] Another essential measure for quantum memories is the storage time. It indicates the time over which the quantum state remains faithfully stored. In particular, for long-distance quantum communication memories, the communication time between distant nodes imposes a lower boundary on the quantum memory storage time.[11] Furthermore, the bandwidth of the quantum memory determines the wavelength compatibility with the interfacing entangled photon source and affects the achievable repetition rates.[45]

Several quantum memory schemes have been developed so far. Quantum memories based on EIT were first described by Fleischhauer and Lukin in 2002.[135] Since then this scheme was also demonstrated beyond warm[38,135] and cold atomic vapors,[136] also in Bose-Einstein condensates[137] or rare-earth doped solid-state materials[138]. EIT is an optical phenomenon in atoms that uses quantum interference to induce transparency into an otherwise resonant and opaque medium and can consequently slow down or store light with storage times > 1 s for quantum memories.[139] EIT based memories have shown record high storage efficiencies of 92%,[140] fidelities close to unity for single photon polarization qubits,[141] as well as very recently entanglement transfer between photon and memory[142]. However, the EIT resonance is typically very small in the order of few MHz, resulting in 3 orders of magnitudes of mismatch in bandwidth with QD photons. Major advances for atomic vapor-based quantum memories have been achieved using off-resonant interactions, such as the gradient-echo spin-wave memory (GEM). The first off-resonant Raman memories already combined both high efficiency (30%) and large spectral bandwidth (1.5 GHz).[143] Since then, the memory's efficiency was increased up to 87%[144] and its storage time up to milliseconds[145]. Major efforts have been taken to bring the average fidelity up to 72% for polarization qubits with storage lifetimes of about 20 μs[146] and around 90% for storage time of 700 ns[147]. While GEM memories, a protocol based on controlled reversible inhomogeneous broadening, reach high efficiencies such as 87 % after a storage time of 3.7 μs in atomic vapor,[144] and 69 % after a storage time of 1.3 μs in rare-earth doped crystals[148] for modulated laser light, there is a trade-off between efficiency and bandwidth, since the bandwidth is proportional to the broadening of the atomic ensemble prior to absorption of the write pulse. In contrast, the atomic frequency comb storage (AFC) protocol can, combined with an optical cavity, achieve high efficiency (theoretically up to 95 %,[149] experimentally 56 %[150]) without degrading the bandwidth. The AFC storage protocol stores light at the single-photon level[151], but on-demand readout requires additional efforts[152]. To store one photon per pulse onto an ensemble of naturally trapped atoms in a solid, a light field has to be coherently absorbed in a suitably prepared solid-state atomic medium.[153] First implementations of storing weak laser pulses with less than one photon per pulse in a prepared solid-state atomic ensemble, that is naturally trapped in a solid, store the state of light for a pre-determined time of up to 1 μs.[154] The coherence of the process is verified by the performance of an interference experiment with two stored weak pulses. Visibilities above 95% were achieved after the subtraction of detector dark counts. The multimode character of AFC memories is important for future quantum repeater applications, speeding up the repeater protocol. Furthermore, the large selection of suitable rare-earth ion transitions makes these types of memory very flexible with respect of operation wavelength. One very recent contender for an efficient rare-earth doped crystal quantum memory is $^{167}$Er:Y$_2$SiO$_5$. It has comparable optical depth and optical pumping efficiency as Pr:Y$_2$SiO$_5$, which is often used as a quantum memory. In addition, $^{167}$Erbium-based quantum memory can operate at 1538 nm, possible with a larger memory bandwidth and reduced noise.[155]

Much effort has been put into improving storage times and efficiencies of quantum memories to enable long-distance communication. However, research up to now has devoted less attention to building quantum memories with zero noise output, which can render the memory classical by destroying the quantum characteristics of the stored light. A noise-free and on-demand atomic-frequency comb-based protocol is proposed by Horvath et al..[156] Utilizing the Stark effect to the established atomic frequency comb protocol they could demonstrate a recall efficiency of 38 % for a storage time of 0.8 μs in a solid-state system.[156] Ultra-low noise operation was also realized in warm atomic vapor by deploying destructive interference of dark-state polaritons using an auxiliary field in an EIT protocol.[147] Another noise-free quantum memory protocol is based on two-photon off-resonant cascaded absorption (ORCA).[157] This ORCA protocol provides a viable real-world platform that does not measurably degrade the quantum character of the recalled light compared to the input. By using this protocol, and the less detuned variant called FLAME,[158] the successful storage of GHz-bandwidth heralded single photons in a warm atomic vapor with no added noise has been demonstrated. Thus, meeting the stringent noise requirements for quantum memories with high-speed and room-temperature operation, though, with short storage times.

The requirements imposed on the quantum memory and the need for various memory schemes emphasize the complexity of obtaining efficient interfacing between a quantum light source and a quantum memory. For quantum repeaters relying on QDs many challenges have to be addressed. One of them is the wavelength matching of the source with the memory. **Table 1** summarizes frequency-matched memories and QD pairs. Thanks to the semiconductor platform and post-growth fine-tuning techniques (i.e. strain, electric or magnetic field) QDs have a large range of operational wavelengths. Allowing to be frequency matched- to several different types of quantum memories. However, the bandwidth mismatch between quantum memory and quantum dot photons is still a major challenge, hindering the realization of an efficient interface. With the recent advent of off-resonant ensemble-based quantum memory schemes such as Raman, ORCA, and FORD[159] the bandwidth of the memory can be increased. This comes with the cost of either increased intrinsic noise or reduced storage time.

In comparison, atomic frequency comb based rare earth quantum memories can take advantage of the inhomogeneous broadening of the dopants to generate a broader frequency comb. So far AFC memories have been the only memory type which successfully stored quantum dot photons.[160] Thus, thulium dopes Y$_3$Al$_5$O$_{12}$ crystal have recently become a strong contender for hybrid quantum repeaters, since these memories could be frequency matched with the state-of-the-art GaAs quantum dots. Tm:Y$_3$Al$_5$O$_{12}$ crystals already offer absorption of 90% of input photons and a memory efficiency of 27.5% over a 500 MHz bandwidth.[161]

| Quantum memory platform | Quantum memory wavelength regime [nm] | Material of frequency matched quantum dot |
|---|---|---|
| Rubidium | 780 and 795 | GaAs |
|  | 1529 | InAs and InAsP |
| Cesium | 852 and 895 | InGaAs and InAsP |
| $^{167}$Er: Y$_2$SiO$_5$ | 1538 | InAs and InAsP |
| Tm: Y$_3$Al$_5$O$_{12}$ | 793 | GaAs |
| Nd$^{3+}$: YVO$_4$ | 880 | InGaAs and InAsP |

*Table1: Promising hybrid systems with QDs and quantum memories*

In the following we will briefly give an overview of experimental efforts taken to interface quantum dots with possible quantum memory platforms. In 2011, the first successful attempt to interface photons from QDs with warm rubidium vapor was performed by Akopian et al.,[162] who demonstrated slowing down of single photons when travelling through the warm rubidium vapor. The GaAs QDs were designed to emit at around 780 nm, and an external magnetic field was used to fine-tune the exciton to the $^{87}$Rb D$_2$ transitions. Later in 2017, Huang et al.[163] extended this results to QD-LED integrated onto piezoelectric actuators, thus showing that both the excitation and the wavelength-tunability can be provided on-chip. The extension of this approach to entangled photons requires a different level of complexity, as energy tuning has to be achieved without affecting the degree of entanglement of the source, i.e., without opening the FSS. As mentioned before, this task can be accomplished using a micro-machined piezoelectric actuator. Trotta *et al.*[72] could demonstrate with In(Ga)As QDs that strain-engineering can be used to tune the FSS to zero and simultaneously in resonance with the D$_1$ lines of cesium. Beside slow entangled photons, they could confirm that the use of atomic vapor does not affect the degree of entanglement of the sources. All the experiments discussed so far, as well as additional studies on that topic[164,165], highlight that current technology is ready to interface artificial and natural atomic systems. However, these approaches focused only on slow single/entangled photons while quantum repeater schemes demand efficient storage, with long storing times,[166] as well as the retrieval of photons from the storing media,[23,167] i.e., a full quantum memory for QD photons. Recent experimental efforts[168] have shown that one possibility is based on atomic gases. Here, various approaches towards a quantum memory can be used, as for example Raman transitions[129] and electromagnetically induced transparency (EIT)[169]. Wolters *et al.*[168] were able to demonstrate a memory in warm Rb vapor with on-demand storage and retrieval via a near-resonant scheme based on EIT using attenuated laser pulses that are comparable to the bandwidth and pulse intensity of photons from QDs. For a storage time of 50 ns, they could demonstrate an intrinsic memory efficiency of $v_{int}$ = 17 % and an end-to-end efficiency of the fiber-coupled memory of $v^{50ns}_{e2e}$= 5 %. This end-to-end efficiency can be boosted to around 35% by primarily optimizing the filtering systems and using control pulses with higher power. Additionally, longer storage times can be achieved by the use of a cold atomic system.[170] Other approaches demonstrated by Meyer *et al.*[171] show a direct coupling between QDs and trapped ions.

## 2.6 Entanglement swapping

After introducing the main hardware elements required for a memory-based quantum repeater, namely the entangled photon source and the quantum memory, we now move on to discuss the basic protocol used to operate its elementary blocks, that is entanglement swapping, and its experimental realization.

In the recent years, efforts in the research community working on QDs-based entangled photon sources have led to the demonstration of quantum protocols which rely on entanglement and Bell state measurements, targeting application in long-distance communication. Such concept was first introduced with the conception of quantum teleportation. Quantum teleportation is the transmission of a quantum state of a particle to another distant one without the transmission of the particle itself. After its theoretical proposal by Bennett *et al.*[172] in 1993, the first experimental proofs were achieved by Bouwmeester *et al.*[30] in 1997 and Boschi *et al.*[29] in 1998. With entangled photons produced by parametric down-conversion the polarization state from one photon could be successfully teleported to another.

The first quantum teleportation experiment using an electrically-driven entangled photon source based on an InGaAs QD was performed by Nilsson *et al.*[173] in 2013. The researchers managed to transfer the polarization state of a photon from a QD onto one of the photons of an entangled pair emitted from the same QD with a short time delay. They achieved an average teleportation fidelity of 0.704 using continuous pumping and temporal filtering. To move towards a deterministic operation scheme, we have to wait for Reindl *et al.*[87] in 2018. This time the source is used in pulsed mode under two-photon resonant optical excitation, without temporal postelection on the relevant coincidence events, background subtraction, or postprocessing of the measured data, achieving an averaged teleportation fidelity of 0.75(2). The teleportation fidelity, that is estimated from the analysis of three-fold coincidences, is dependent on two main parameters: first, the HOM visibility of the two exciton photons, and second, the initial entanglement fidelity to the expected Bell state. Consequently, imperfections of the source decrease the HOM visibility and thus significantly lower the protocol fidelity.

The role of specific QD imperfections on the performance of quantum teleportation was recently investigated in an experimental study by Basso Basset *et al.*.[123] The researchers investigate the effect of the protocol implementations on the teleportation fidelity. By only adjusting the Bell state measurement apparatus and applying moderate spectral filtering they could demonstrate that the average teleportation fidelity of a QD with below average figures of merit, i.e., entanglement and indistinguishability, can be brought from below the classical limit to values as high as 0.842(14). This indicates that even though the FSS and two-photon interference visibility may not be optimal in as-grown QDs, their impact on the fidelity of the teleportation can be partially mitigated, with the cost of brightness.

The natural extension of these quantum teleportation experiments has led to recent demonstration of entanglement swapping with photons from QDs. Using a protocol that is analogous to the seminal experiment performed by Pan *et al.*[27] in 1998, Basso Basset *et al.*[174] in 2019 performed the protocol with GaAs QDs without recurring to spectral or temporal filtering and using resonant two-photon optical excitation for the deterministic generation of entangled photons. The scheme, previously presented in the introduction section, corresponds to the elementary block of the quantum repeater illustrated in **Figure 1a**, minus the presence of the quantum memories. After two entangled photon pairs are generated in the state of polarization φ⁺, one photon from each pair is sent to a BSM apparatus, which can perform a projective measurement on the state ψ⁻. Successful BSM establishes entanglement between the remaining previously uncorrelated photons. The state of these photons was reconstructed in the experiment by means of a full quantum state tomography, resulting in the density matrix reported in **Figure 6a**. The fidelity of the density matrix to the expected Bell state ψ⁻ is 0.58(4), which is above the classical limit. This value is mainly attributed to the achieved HOM visibility of 0.63(2). A notable improvement of the swapping fidelity is expected by implementing a polarization-selective BSM with increased accuracy, as demonstrated—and previously discussed—for quantum teleportation with imperfect sources.[123] This point is better showcased in the simulation in **Figure 6b**. It describes the dependence of the swapping fidelity on the most relevant figures of merit of the QD, namely the ratio between FSS and natural linewidth and the HOM visibility,[175] for a polarization-selective BSM. In particular, the black dot marks the expected swapping fidelity of approximately 0.77 for the QD used in the previously discussed experiment. Similar results were achieved by Zopf et al,[176] but using temporal post-selection of the emitted photons.

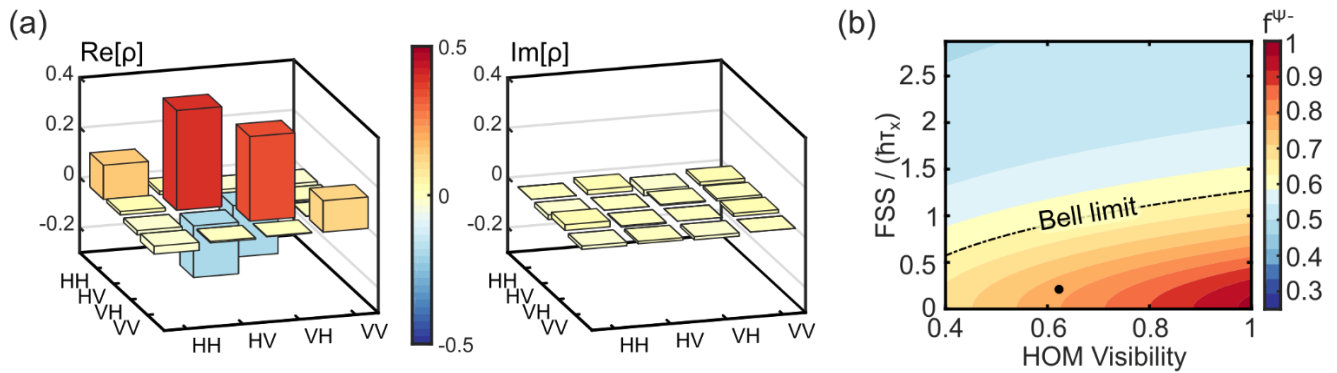

*Figure 6: Entanglement swapping. a) Two-photon polarization density matrix experimentally obtained using quantum state tomography. Measurement results from an entanglement swapping operation performed on polarization-entangled photons emitted from a QD and triggered by a Bell state measurement (BSM) relying on two-photon interference at a beam splitter. Adapted from Ref.[174] b) Contour plot of the simulated fidelity of entanglement swapping as a function of Hong-Ou-Mandel (HOM) visibility and the ratio between fine structure splitting (FSS) and natural linewidth, under the assumption of a polarization-selective BSM. The expected improvement for this setup variation on the QD measured in panel (a) is estimated and indicated by the black dot. The threshold for the violation of Bell's inequality is marked by the dash-dotted line.*

## 3. Discussion

Entangled photon sources based on QDs have faced and continue facing several challenges in their way towards application in quantum repeaters. The first to be addressed has been the FSS caused by the asymmetries in the confinement potential of the QDs. Even though fabrication protocols are steadily improving, the control of the process is still not impeccable, leading to anisotropies in the QDs and limiting the degree of entanglement achievable in absence of post-selection strategies. However, it has been demonstrated that the FSS can be conveniently suppressed by applying external perturbations, allowing high entanglement fidelities with on-demand emission.[36] The advances in the entanglement and state preparation figures of merit of the source lead to the achievement of seminal experiments, such as quantum teleportation[87] and entanglement swapping[174,176] that demonstrate the potential of entangled photons from QDs for quantum repeaters.

Nonetheless, to overcome the proof-of-concept stage of the current applications of QD in quantum information experiments, first, the brightness of the source needs to be improved—a major obstacle for quantum emitters embedded in a solid-state matrix. Extraction efficiencies around 10% are typical for the previously cited protocol realizations and, thus, far off from the desired values. A decisive step forward is based on the exploitation of light-matter interaction, with circular Bragg gratings being a promising approach to push the pair-extraction efficiency up to near-unity values, currently up to 65%.[177] Additionally, circular Bragg grating cavities allow, via the Purcell effect, to tackle the issue related to the limited indistinguishability of photons emitted in a radiative cascade. It has been shown that the biexciton-exciton cascaded process introduces a time correlation depending on the ratio of the biexciton and exciton lifetimes. This inherently limits the degree of indistinguishability for both emitted photons.[112] Selective Purcell enhancement together with mild frequency filtering could alleviate this hurdle, as suggested in a recent theoretical study.[178]

The aforementioned experiments, quantum teleportation and entanglement swapping with entangled photons from QDs, were both conducted using the same source, which for some real-life applications in communication, namely distant quantum repeaters, is no option. In a quantum network the goal is to generate entanglement between two distant quantum repeaters, so performing entanglement swapping between photon pairs from two different QDs. The highest reported two-photon interference visibility of V = 0.51(5) discussed above[119] points out the indispensable need for source optimization. A major contribution to the low remote visibility results from the fabrication methods that currently do not reproducibly provide emitters with Fourier-limited linewidth. Nonetheless, a powerful approach based on charge control via an applied electric field[179] has already been applied in single-photon sources, showcasing the viability of mitigation strategies. Additionally, faster recombination times due to Purcell enhancement can also be beneficial, provided that further causes for spectral wandering are not introduced in the microfabrication process.

In the race for the construction of a QD-based quantum repeater, another major challenge is to find a suitable memory for QD photons. In this direction, the first steps to build up an artificial-natural atomic interface have been moved by the demonstration of slow single photons[162] as well as slow entangled photons[72] due to the interface of the QD photons with atomic vapors. Even though these experiments investigated only a feature preliminary to the functionalities of a quantum repeater, they provide an encouraging basis for quantum-memory experiments based on atomic vapors; especially, quantum memories based on EIT or GEM. The state-of-the-art intrinsic memory efficiency of 17%[168] using an EIT-based scheme interfaced with a QD can be possibly improved by using a GEM scheme, which reaches efficiencies up to 87%[144] for weak laser pulses. However, in the GEM scheme, high efficiencies compromise the bandwidth, which might not be compatible with a QD-based entangled photon source. Therefore, other memory schemes such as AFC, that could theoretically reach an efficiency up to 95%[149] without the tradeoff of bandwidth, might be attractive for the future. However, improving the efficiency of quantum memories alone most likely will not suffice, because noise added by the memory can still render the memory classical by destroying the quantum characteristic of the stored light. One solution might provide recent developments on noise-free protocols, i.e. the ORCA protocol that is based on room-temperature atomic vapor,[157] or the zero-noise output protocol for AFC[156]. These protocols fill a missing gap allowing possible future applicability of memory-based quantum repeater schemes.

For the overall goal of using this technology in the already existing fiber structure, the wavelength has to be matched to the telecom C-band, where the attenuation in silica fibers has a minimum. A possible solution, however, with the cost of efficiency, provide frequency down-conversion techniques to adapt the emission wavelength.[117] As an alternative, all of the general concepts discussed can also be adapted to different material systems, such as InGaAs QDs, whose emission wavelength can be extended to the telecom C-band.[53,103,180–182] For this purpose, different memory systems for QDs emitting in the telecom C-band are developed.[155,183,184]

In conclusion, combining the state-of-the-art QDs with photonic cavities, to maximize the extraction efficiency, and with the well-established tuning techniques, such as electrical or strain tuning to erase FSS and adjust the emission wavelength of the entangled photon pair, all the desirable features for an efficient entangled photon source are at hand: negligible multiphoton emission,[34] high extraction efficiency,[21] degree of entanglement,[20] photon-indistinguishability,[35] wavelength-tunability,[72] with on-demand generation[37]. This rapid progress in the field has recently led to out-of-the-lab demonstrations of entanglement-based quantum key distribution[185,186] and further applications in quantum communication are to be envisaged. The high compatibility with some quantum memories make QDs attractive for quantum repeaters, but their realization is evidently a grand challenge and certainly requires extreme efforts. However, the building blocks have been set up and the dream of a solid-state-based quantum network is expected to become a reality in the foreseeable future.

Acknowledgements
This work was financially supported by the European Research Council (ERC) under the European Union's Horizon 2020 Research and Innovation Programme (SPQRel, grant agreement no. 679183), by MIUR (Ministero dell'Istruzione, dellUniversita e della Ricerca) via project PRIN 2017 Taming complexity via Quantum Strategies a Hybrid Integrated Photonic approach (QUSHIP) Id. 2017SRNBRK, by the European Union's Horizon 2020 research and innovation program under Grant Agreement no. 820423 (S2QUIP) and No. 899814 (Qurope).

Bibliography

[1] H. J. Kimble, *Nature* **2008**, *453*, 1023.
[2] S. Wehner, D. Elkouss, R. Hanson, *Science* **2018**, *362*, DOI 10.1126/science.aam9288.
[3] V. Scarani, C. Kurtsiefer, *Theoretical Computer Science* **2014**, *560*, 27.
[4] F. Xu, X. Ma, Q. Zhang, H.-K. Lo, J.-W. Pan, *Rev. Mod. Phys.* **2020**, *92*, 025002.
[5] M. Takeoka, S. Guha, M. M. Wilde, *Nature Communications* **2014**, *5*, 5235.
[6] C. Antonelli, M. Shtaif, M. Brodsky, *Phys. Rev. Lett.* **2011**, *106*, 080404.
[7] W. K. Wootters, W. H. Zurek, *Nature* **1982**, *299*, 802.
[8] W. J. Munro, A. M. Stephens, S. J. Devitt, K. A. Harrison, K. Nemoto, *Nature Photonics* **2012**, *6*, 777.
[9] Z.-D. Li, R. Zhang, X.-F. Yin, L.-Z. Liu, Y. Hu, Y.-Q. Fang, Y.-Y. Fei, X. Jiang, J. Zhang, L. Li, N.-L. Liu, F. Xu, Y.-A. Chen, J.-W. Pan, *Nature Photonics* **2019**, DOI 10.1038/s41566-019-0468-5.
[10] J. Borregaard, H. Pichler, T. Schröder, M. D. Lukin, P. Lodahl, A. S. Sørensen, *Phys. Rev. X* **2020**, *10*, 021071.
[11] N. Sangouard, C. Simon, H. de Riedmatten, N. Gisin, *Rev. Mod. Phys.* **2011**, *83*, 33.
[12] S. Lloyd, M. S. Shahriar, J. H. Shapiro, P. R. Hemmer, *Physical Review Letters* **2001**, *87*, DOI 10.1103/PhysRevLett.87.167903.
[13] P. van Loock, W. Alt, C. Becher, O. Benson, H. Boche, C. Deppe, J. Eschner, S. Höfling, D. Meschede, P. Michler, F. Schmidt, H. Weinfurter, *Advanced Quantum Technologies* **2020**, *3*, 1900141.
[14] J. Borregaard, A. S. Sørensen, P. Lodahl, *Advanced Quantum Technologies* **2019**, *2*, 1800091.
[15] R. Stockill, C. Le Gall, C. Matthiesen, L. Huthmacher, E. Clarke, M. Hugues, M. Atatüre, *Nature Communications* **2016**, *7*, 12745.
[16] D. Cogan, O. Kenneth, N. H. Lindner, G. Peniakov, C. Hopfmann, D. Dalacu, P. J. Poole, P. Hawrylak, D. Gershoni, *Phys. Rev. X* **2018**, *8*, 041050.
[17] K. Azuma, K. Tamaki, H.-K. Lo, *Nature Communications* **2015**, *6*, DOI 10.1038/ncomms7787.
[18] R. Raussendorf, H. J. Briegel, *Phys. Rev. Lett.* **2001**, *86*, 5188.
[19] I. Schwartz, D. Cogan, E. R. Schmidgall, Y. Don, L. Gantz, O. Kenneth, N. H. Lindner, D. Gershoni, *Science* **2016**, *354*, 434.
[20] D. Huber, M. Reindl, S. F. Covre da Silva, C. Schimpf, J. Martín-Sánchez, H. Huang, G. Piredda, J. Edlinger, A. Rastelli, R. Trotta, *Physical Review Letters* **2018**, *121*, DOI 10.1103/PhysRevLett.121.033902.
[21] J. Liu, R. Su, Y. Wei, B. Yao, S. F. C. da Silva, Y. Yu, J. Iles-Smith, K. Srinivasan, A. Rastelli, J. Li, X. Wang, *Nature Nanotechnology* **2019**, *14*, 586.
[22] H. Wang, H. Hu, T.-H. Chung, J. Qin, X. Yang, J.-P. Li, R.-Z. Liu, H.-S. Zhong, Y.-M. He, X. Ding, Y.-H. Deng, Q. Dai, Y.-H. Huo, S. Höfling, C.-Y. Lu, J.-W. Pan, *Physical Review Letters* **2019**, *122*, 113602.
[23] H.-J. Briegel, W. Dür, J. I. Cirac, P. Zoller, *Physical Review Letters* **1998**, *81*, 5932.
[24] M. Żukowski, A. Zeilinger, M. A. Horne, A. K. Ekert, *Physical Review Letters* **1993**, *71*, 4287.
[25] E. Waks, A. Zeevi, Y. Yamamoto, *Phys. Rev. A* **2002**, *65*, 052310.
[26] C. H. Bennett, G. Brassard, S. Popescu, B. Schumacher, J. A. Smolin, W. K. Wootters, *Phys. Rev. Lett.* **1996**, *76*, 722.
[27] J.-W. Pan, D. Bouwmeester, H. Weinfurter, A. Zeilinger, *Physical Review Letters* **1998**, *80*, 3891.
[28] P. G. Kwiat, K. Mattle, H. Weinfurter, A. Zeilinger, A. V. Sergienko, Y. Shih, *Phys. Rev. Lett.* **1995**, *75*, 4337.
[29] D. Boschi, S. Branca, F. De Martini, L. Hardy, S. Popescu, *Physical Review Letters* **1998**, *80*, 1121.
[30] D. Bouwmeester, J.-W. Pan, K. Mattle, K. Mattle, H. Weinfurter, A. Zeilinger, *Nature* **1997**, *390*, 575.

[31]  V. Scarani, H. de Riedmatten, I. Marcikic, H. Zbinden, N. Gisin, *Eur. Phys. J. D* **2005**, *32*, 129.
[32]  O. Benson, C. Santori, M. Pelton, Y. Yamamoto, *Physical Review Letters* **2000**, *84*, 2513.
[33]  L. Hanschke, K. A. Fischer, S. Appel, D. Lukin, J. Wierzbowski, S. Sun, R. Trivedi, J. Vučković, J. J. Finley, K. Müller, *npj Quantum Inf* **2018**, *4*, 43.
[34]  L. Schweickert, K. D. Jöns, K. D. Zeuner, S. F. Covre da Silva, H. Huang, T. Lettner, M. Reindl, J. Zichi, R. Trotta, A. Rastelli, V. Zwiller, *Applied Physics Letters* **2018**, *112*, 093106.
[35]  D. Huber, M. Reindl, Y. Huo, H. Huang, J. S. Wildmann, O. G. Schmidt, A. Rastelli, R. Trotta, *Nature Communications* **2017**, *8*, 15506.
[36]  R. Trotta, J. Martín-Sánchez, I. Daruka, C. Ortix, A. Rastelli, *Phys. Rev. Lett.* **2015**, *114*, 150502.
[37]  M. Müller, S. Bounouar, K. D. Jöns, M. Glässl, P. Michler, *Nature Photonics* **2014**, *8*, 224.
[38]  L.-M. Duan, M. D. Lukin, J. I. Cirac, P. Zoller, *Nature* **2001**, *414*, 413.
[39]  M. G. Raymer, I. A. Walmsley, J. Mostowski, B. Sobolewska, *Phys. Rev. A* **1985**, *32*, 332.
[40]  A. Kuzmich, K. Mølmer, E. S. Polzik, *Phys. Rev. Lett.* **1997**, *79*, 4782.
[41]  M. D. Lukin, S. F. Yelin, M. Fleischhauer, *Phys. Rev. Lett.* **2000**, *84*, 4232.
[42]  D. F. Phillips, A. Fleischhauer, A. Mair, R. L. Walsworth, M. D. Lukin, *Physical Review Letters* **2001**, *86*, 783.
[43]  T. Zhong, J. M. Kindem, J. G. Bartholomew, J. Rochman, I. Craiciu, E. Miyazono, M. Bettinelli, E. Cavalli, V. Verma, S. W. Nam, F. Marsili, M. D. Shaw, A. D. Beyer, A. Faraon, *Science* **2017**, *357*, 1392.
[44]  J. Guo, X. Feng, P. Yang, Z. Yu, L. Q. Chen, C.-H. Yuan, W. Zhang, *Nature Communications* **2019**, *10*, 148.
[45]  G. T. Campbell, K. R. Ferguson, M. J. Sellars, B. C. Buchler, P. K. Lam, *arXiv:1902.04313 [physics, physics:quant-ph]* **2019**.
[46]  M. Bayer, G. Ortner, O. Stern, A. Kuther, A. A. Gorbunov, A. Forchel, P. Hawrylak, S. Fafard, K. Hinzer, T. L. Reinecke, S. N. Walck, J. P. Reithmaier, F. Klopf, F. Schäfer, *Physical Review B* **2002**, *65*, DOI 10.1103/PhysRevB.65.195315.
[47]  G. Juska, E. Murray, V. Dimastrodonato, T. H. Chung, S. T. Moroni, A. Gocalinska, E. Pelucchi, *Journal of Applied Physics* **2015**, *117*, 134302.
[48]  A. Rastelli, M. Stoffel, A. Malachias, T. Merdzhanova, G. Katsaros, K. Kern, T. H. Metzger, O. G. Schmidt, *Nano Lett.* **2008**, *8*, 1404.
[49]  M. Müller, A. Cerezo, G. D. W. Smith, L. Chang, S. S. A. Gerstl, *Appl. Phys. Lett.* **2008**, *92*, 233115.
[50]  G. Bester, S. Nair, A. Zunger, *Phys. Rev. B* **2003**, *67*, 161306.
[51]  F. Basso Basset, S. Bietti, M. Reindl, L. Esposito, A. Fedorov, D. Huber, A. Rastelli, E. Bonera, R. Trotta, S. Sanguinetti, *Nano Letters* **2018**, *18*, 505.
[52]  A. J. Hudson, R. M. Stevenson, A. J. Bennett, R. J. Young, C. A. Nicoll, P. Atkinson, K. Cooper, D. A. Ritchie, A. J. Shields, *Phys. Rev. Lett.* **2007**, *99*, 266802.
[53]  N. Akopian, N. H. Lindner, E. Poem, Y. Berlatzky, J. Avron, D. Gershoni, B. D. Gerardot, P. M. Petroff, *Physical Review Letters* **2006**, *96*, DOI 10.1103/PhysRevLett.96.130501.
[54]  R. M. Stevenson, A. J. Hudson, A. J. Bennett, R. J. Young, C. A. Nicoll, D. A. Ritchie, A. J. Shields, *Phys. Rev. Lett.* **2008**, *101*, 170501.
[55]  A. Fognini, A. Ahmadi, M. Zeeshan, J. T. Fokkens, S. J. Gibson, N. Sherlekar, S. J. Daley, D. Dalacu, P. J. Poole, K. D. Jöns, V. Zwiller, M. E. Reimer, *ACS Photonics* **2019**, *6*, 1656.
[56]  S. Bounouar, G. Rein, K. Barkemeyer, J. Schleibner, P. Schnauber, M. Gschrey, J.-H. Schulze, A. Strittmatter, S. Rodt, A. Knorr, A. Carmele, S. Reitzenstein, *Phys. Rev. B* **2020**, *102*, 045304.
[57]  J. Huwer, R. M. Stevenson, J. Skiba-Szymanska, M. B. Ward, A. J. Shields, M. Felle, I. Farrer, D. A. Ritchie, R. V. Penty, *Phys. Rev. Applied* **2017**, *8*, 024007.
[58]  J. D. Plumhof, R. Trotta, A. Rastelli, O. G. Schmidt, *Nanoscale Research Letters* **2012**, *7*, 336.
[59]  D. Huber, M. Reindl, J. Aberl, A. Rastelli, R. Trotta, *Journal of Optics* **2018**, *20*, 073002.

[60] A. J. Bennett, M. A. Pooley, R. M. Stevenson, M. B. Ward, R. B. Patel, A. B. de la Giroday, N. Sköld, I. Farrer, C. A. Nicoll, D. A. Ritchie, A. J. Shields, *Nature Physics* **2010**, *6*, 947.
[61] B. D. Gerardot, S. Seidl, P. A. Dalgarno, R. J. Warburton, D. Granados, J. M. Garcia, K. Kowalik, O. Krebs, K. Karrai, A. Badolato, P. M. Petroff, *Appl. Phys. Lett.* **2007**, *90*, 041101.
[62] M. M. Vogel, S. M. Ulrich, R. Hafenbrak, P. Michler, L. Wang, A. Rastelli, O. G. Schmidt, *Appl. Phys. Lett.* **2007**, *91*, 051904.
[63] J. D. Plumhof, V. Křápek, F. Ding, K. D. Jöns, R. Hafenbrak, P. Klenovský, A. Herklotz, K. Dörr, P. Michler, A. Rastelli, O. G. Schmidt, *Phys. Rev. B* **2011**, *83*, 121302.
[64] S. Seidl, M. Kroner, A. Högele, K. Karrai, R. J. Warburton, A. Badolato, P. M. Petroff, *Appl. Phys. Lett.* **2006**, *88*, 203113.
[65] C. E. Kuklewicz, R. N. E. Malein, P. M. Petroff, B. D. Gerardot, *Nano Lett.* **2012**, *12*, 3761.
[66] R. Singh, G. Bester, *Phys. Rev. Lett.* **2010**, *104*, 196803.
[67] M. Gong, W. Zhang, G.-C. Guo, L. He, *Phys. Rev. Lett.* **2011**, *106*, 227401.
[68] R. Trotta, E. Zallo, C. Ortix, P. Atkinson, J. D. Plumhof, J. van den Brink, A. Rastelli, O. G. Schmidt, *Physical Review Letters* **2012**, *109*, DOI 10.1103/PhysRevLett.109.147401.
[69] R. Trotta, J. S. Wildmann, E. Zallo, O. G. Schmidt, A. Rastelli, *Nano Letters* **2014**, *14*, 3439.
[70] J. Wang, M. Gong, G.-C. Guo, L. He, *Phys. Rev. Lett.* **2015**, *115*, 067401.
[71] Y. Chen, J. Zhang, M. Zopf, K. Jung, Y. Zhang, R. Keil, F. Ding, O. G. Schmidt, *Nature Communications* **2016**, *7*, 10387.
[72] R. Trotta, J. Martín-Sánchez, J. S. Wildmann, G. Piredda, M. Reindl, C. Schimpf, E. Zallo, S. Stroj, J. Edlinger, A. Rastelli, *Nature Communications* **2016**, *7*, 10375.
[73] C. L. Salter, R. M. Stevenson, I. Farrer, C. A. Nicoll, D. A. Ritchie, A. J. Shields, *Nature* **2010**, *465*, 594.
[74] R. M. Stevenson, C. L. Salter, J. Nilsson, A. J. Bennett, M. B. Ward, I. Farrer, D. A. Ritchie, A. J. Shields, *Phys. Rev. Lett.* **2012**, *108*, 040503.
[75] R. Trotta, P. Atkinson, J. D. Plumhof, E. Zallo, R. O. Rezaev, S. Kumar, S. Baunack, J. R. Schröter, A. Rastelli, O. G. Schmidt, *Advanced Materials* **2012**, *24*, 2668.
[76] J. R. A. Müller, R. M. Stevenson, J. Skiba-Szymanska, G. Shooter, J. Huwer, I. Farrer, D. A. Ritchie, A. J. Shields, *Phys. Rev. Research* **2020**, *2*, 043292.
[77] J. Zhang, J. S. Wildmann, F. Ding, R. Trotta, Y. Huo, E. Zallo, D. Huber, A. Rastelli, O. G. Schmidt, *Nature Communications* **2015**, *6*, DOI 10.1038/ncomms10067.
[78] S. Stufler, P. Machnikowski, P. Ester, M. Bichler, V. M. Axt, T. Kuhn, A. Zrenner, *Physical Review B* **2006**, *73*, DOI 10.1103/PhysRevB.73.125304.
[79] K. Brunner, G. Abstreiter, G. Böhm, G. Tränkle, G. Weimann, *Phys. Rev. Lett.* **1994**, *73*, 1138.
[80] H. Jayakumar, A. Predojević, T. Huber, T. Kauten, G. S. Solomon, G. Weihs, *Phys. Rev. Lett.* **2013**, *110*, 135505.
[81] C. Hopfmann, W. Nie, N. L. Sharma, C. Weigelt, F. Ding, O. G. Schmidt, *Phys. Rev. B* **2021**, *103*, 075413.
[82] S. Bounouar, M. Müller, A. M. Barth, M. Glässl, V. M. Axt, P. Michler, *Phys. Rev. B* **2015**, *91*, 161302.
[83] P. Senellart, G. Solomon, A. White, *Nature Nanotechnology* **2017**, *12*, 1026.
[84] H. Benisty, H. De Neve, C. Weisbuch, *IEEE Journal of Quantum Electronics* **1998**, *34*, 1612.
[85] W. L. Barnes, G. Björk, J. M. Gérard, P. Jonsson, J. a. E. Wasey, P. T. Worthing, V. Zwiller, *Eur. Phys. J. D* **2002**, *18*, 197.
[86] Y. Chen, M. Zopf, R. Keil, F. Ding, O. G. Schmidt, *Nature Communications* **2018**, *9*, 2994.
[87] M. Reindl, D. Huber, C. Schimpf, S. F. C. da Silva, M. B. Rota, H. Huang, V. Zwiller, K. D. Jöns, A. Rastelli, R. Trotta, *Science Advances* **2018**, *4*, eaau1255.

[88]  O. Gazzano, S. Michaelis de Vasconcellos, C. Arnold, A. Nowak, E. Galopin, I. Sagnes, L. Lanco, A. Lemaître, P. Senellart, *Nature Communications* **2013**, *4*, DOI 10.1038/ncomms2434.
[89]  N. Tomm, A. Javadi, N. O. Antoniadis, D. Najer, M. C. Löbl, A. R. Korsch, R. Schott, S. R. Valentin, A. D. Wieck, A. Ludwig, R. J. Warburton, *Nature Nanotechnology* **2021**, 1.
[90]  A. Dousse, J. Suffczyński, A. Beveratos, O. Krebs, A. Lemaître, I. Sagnes, J. Bloch, P. Voisin, P. Senellart, *Nature* **2010**, *466*, 217.
[91]  M. A. M. Versteegh, M. E. Reimer, K. D. Jöns, D. Dalacu, P. J. Poole, A. Gulinatti, A. Giudice, V. Zwiller, *Nature Communications* **2014**, *5*, 5298.
[92]  L. Sapienza, M. Davanço, A. Badolato, K. Srinivasan, *Nature Communications* **2015**, *6*, 7833.
[93]  M. Moczała-Dusanowska, Ł. Dusanowski, O. Iff, T. Huber, S. Kuhn, T. Czyszanowski, C. Schneider, S. Höfling, *ACS Photonics* **2020**, *7*, 3474.
[94]  N. Lütkenhaus, J. Calsamiglia, K.-A. Suominen, *Phys. Rev. A* **1999**, *59*, 3295.
[95]  Y. H. Shih, C. O. Alley, *Phys. Rev. Lett.* **1988**, *61*, 2921.
[96]  E. Knill, R. Laflamme, G. J. Milburn, *Nature* **2001**, *409*, 46.
[97]  H. Wang, J. Qin, X. Ding, M.-C. Chen, S. Chen, X. You, Y.-M. He, X. Jiang, L. You, Z. Wang, C. Schneider, J. J. Renema, S. Höfling, C.-Y. Lu, J.-W. Pan, *Phys. Rev. Lett.* **2019**, *123*, 250503.
[98]  C. K. Hong, Z. Y. Ou, L. Mandel, *Physical Review Letters* **1987**, *59*, 2044.
[99]  B. Kambs, C. Becher, *New J. Phys.* **2018**, *20*, 115003.
[100] A. Kiraz, M. Atatüre, A. Imamoğlu, *Phys. Rev. A* **2004**, *69*, 032305.
[101] E. V. Denning, J. Iles-Smith, N. Gregersen, J. Mork, *Opt. Mater. Express, OME* **2020**, *10*, 222.
[102] A. V. Kuhlmann, J. Houel, A. Ludwig, L. Greuter, D. Reuter, A. D. Wieck, M. Poggio, R. J. Warburton, *Nature Phys* **2013**, *9*, 570.
[103] H. Vural, S. L. Portalupi, P. Michler, *Appl. Phys. Lett.* **2020**, *117*, 030501.
[104] A. J. Bennett, D. C. Unitt, A. J. Shields, P. Atkinson, D. A. Ritchie, *Opt. Express, OE* **2005**, *13*, 7772.
[105] S. Ates, S. M. Ulrich, S. Reitzenstein, A. Löffler, A. Forchel, P. Michler, *Phys. Rev. Lett.* **2009**, *103*, 167402.
[106] L. Besombes, K. Kheng, L. Marsal, H. Mariette, *Phys. Rev. B* **2001**, *63*, 155307.
[107] A. Reigue, J. Iles-Smith, F. Lux, L. Monniello, M. Bernard, F. Margaillan, A. Lemaitre, A. Martinez, D. P. S. McCutcheon, J. Mørk, R. Hostein, V. Voliotis, *Phys. Rev. Lett.* **2017**, *118*, 233602.
[108] J. Iles-Smith, D. P. S. McCutcheon, A. Nazir, J. Mørk, *Nature Photonics* **2017**, *11*, 521.
[109] A. Thoma, P. Schnauber, M. Gschrey, M. Seifried, J. Wolters, J.-H. Schulze, A. Strittmatter, S. Rodt, A. Carmele, A. Knorr, T. Heindel, S. Reitzenstein, *Phys. Rev. Lett.* **2016**, *116*, 033601.
[110] C. Schimpf, M. Reindl, P. Klenovský, T. Fromherz, S. F. Covre Da Silva, J. Hofer, C. Schneider, S. Höfling, R. Trotta, A. Rastelli, *Optics Express* **2019**, *27*, 35290.
[111] E. Schöll, L. Hanschke, L. Schweickert, K. D. Zeuner, M. Reindl, S. F. Covre da Silva, T. Lettner, R. Trotta, J. J. Finley, K. Müller, A. Rastelli, V. Zwiller, K. D. Jöns, *Nano Lett.* **2019**, *19*, 2404.
[112] E. Schöll, L. Schweickert, L. Hanschke, K. D. Zeuner, F. Sbresny, T. Lettner, R. Trivedi, M. Reindl, S. F. Covre da Silva, R. Trotta, J. J. Finley, J. Vučković, K. Müller, A. Rastelli, V. Zwiller, K. D. Jöns, *Phys. Rev. Lett.* **2020**, *125*, 233605.
[113] E. B. Flagg, A. Muller, S. V. Polyakov, A. Ling, A. Migdall, G. S. Solomon, *Phys. Rev. Lett.* **2010**, *104*, 137401.
[114] P. Gold, A. Thoma, S. Maier, S. Reitzenstein, C. Schneider, S. Höfling, M. Kamp, *Phys. Rev. B* **2014**, *89*, 035313.
[115] V. Giesz, S. L. Portalupi, T. Grange, C. Antón, L. De Santis, J. Demory, N. Somaschi, I. Sagnes, A. Lemaître, L. Lanco, A. Auffèves, P. Senellart, *Phys. Rev. B* **2015**, *92*, 161302.
[116] M. Zopf, T. Macha, R. Keil, E. Uruñuela, Y. Chen, W. Alt, L. Ratschbacher, F. Ding, D. Meschede, O. G. Schmidt, *Phys. Rev. B* **2018**, *98*, 161302.

[117] J. H. Weber, B. Kambs, J. Kettler, S. Kern, J. Maisch, H. Vural, M. Jetter, S. L. Portalupi, C. Becher, P. Michler, *Nature Nanotechnology* **2019**, *14*, 23.
[118] A. Thoma, P. Schnauber, J. Böhm, M. Gschrey, J.-H. Schulze, A. Strittmatter, S. Rodt, T. Heindel, S. Reitzenstein, *Appl. Phys. Lett.* **2017**, *110*, 011104.
[119] M. Reindl, K. D. Jöns, D. Huber, C. Schimpf, Y. Huo, V. Zwiller, A. Rastelli, R. Trotta, *Nano Lett.* **2017**, *17*, 4090.
[120] H. Wang, Z.-C. Duan, Y.-H. Li, S. Chen, J.-P. Li, Y.-M. He, M.-C. Chen, Y. He, X. Ding, C.-Z. Peng, C. Schneider, M. Kamp, S. Höfling, C.-Y. Lu, J.-W. Pan, *Phys. Rev. Lett.* **2016**, *116*, 213601.
[121] J. C. Loredo, N. A. Zakaria, N. Somaschi, C. Anton, L. de Santis, V. Giesz, T. Grange, M. A. Broome, O. Gazzano, G. Coppola, I. Sagnes, A. Lemaitre, A. Auffeves, P. Senellart, M. P. Almeida, A. G. White, *Optica, OPTICA* **2016**, *3*, 433.
[122] R. Stockill, M. J. Stanley, L. Huthmacher, E. Clarke, M. Hugues, A. J. Miller, C. Matthiesen, C. Le Gall, M. Atatüre, *Phys. Rev. Lett.* **2017**, *119*, 010503.
[123] F. Basso Basset, F. Salusti, L. Schweickert, M. B. Rota, D. Tedeschi, S. F. Covre da Silva, E. Roccia, V. Zwiller, K. D. Jöns, A. Rastelli, R. Trotta, *npj Quantum Information* **2021**, *7*, 1.
[124] R. M. Stevenson, J. Nilsson, A. J. Bennett, J. Skiba-Szymanska, I. Farrer, D. A. Ritchie, A. J. Shields, *Nature Communications* **2013**, *4*, 2859.
[125] R. M. Macfarlane, *Journal of Luminescence* **2002**, *100*, 1.
[126] A. Batalov, C. Zierl, T. Gaebel, P. Neumann, I.-Y. Chan, G. Balasubramanian, P. R. Hemmer, F. Jelezko, J. Wrachtrup, *Phys. Rev. Lett.* **2008**, *100*, 077401.
[127] R. J. Young, S. J. Dewhurst, R. M. Stevenson, P. Atkinson, A. J. Bennett, M. B. Ward, K. Cooper, D. A. Ritchie, A. J. Shields, *New J. Phys.* **2007**, *9*, 365.
[128] W. Rosenfeld, F. Hocke, F. Henkel, M. Krug, J. Volz, M. Weber, H. Weinfurter, *Phys. Rev. Lett.* **2008**, *101*, 260403.
[129] A. E. Kozhekin, K. Mølmer, E. Polzik, *Phys. Rev. A* **2000**, *62*, 033809.
[130] P. J. Windpassinger, D. Oblak, P. G. Petrov, M. Kubasik, M. Saffman, C. L. G. Alzar, J. Appel, J. H. Müller, N. Kjærgaard, E. S. Polzik, *Phys. Rev. Lett.* **2008**, *100*, 103601.
[131] A. V. Gorshkov, A. André, M. D. Lukin, A. S. Sørensen, *Phys. Rev. A* **2007**, *76*, 033805.
[132] L. A Williamson, J. J Longdell, *New J. Phys.* **2014**, *16*, 073046.
[133] D. Bruss, A. Ekert, C. Macchiavello, *Phys. Rev. Lett.* **1998**, *81*, 2598.
[134] F. Grosshans, P. Grangier, *Phys. Rev. A* **2001**, *64*, 010301.
[135] M. Fleischhauer, M. D. Lukin, *Phys. Rev. A* **2002**, *65*, 022314.
[136] J. Geng, G. T. Campbell, J. Bernu, D. B. Higginbottom, B. M. Sparkes, S. M. Assad, W. P. Zhang, N. P. Robins, P. K. Lam, B. C. Buchler, *New J. Phys.* **2014**, *16*, 113053.
[137] V. Ahufinger, R. Corbalán, F. Cataliotti, S. Burger, F. Minardi, C. Fort, *Optics Communications* **2002**, *211*, 159.
[138] H. Q. Fan, K. H. Kagalwala, S. V. Polyakov, A. L. Migdall, E. A. Goldschmidt, *Phys. Rev. A* **2019**, *99*, 053821.
[139] R. Zhang, S. R. Garner, L. V. Hau, *Phys. Rev. Lett.* **2009**, *103*, 233602.
[140] Y.-F. Hsiao, P.-J. Tsai, H.-S. Chen, S.-X. Lin, C.-C. Hung, C.-H. Lee, Y.-H. Chen, Y.-F. Chen, I. A. Yu, Y.-C. Chen, *Phys. Rev. Lett.* **2018**, *120*, 183602.
[141] Y. Wang, J. Li, S. Zhang, K. Su, Y. Zhou, K. Liao, S. Du, H. Yan, S.-L. Zhu, *Nature Photonics* **2019**, *13*, 346.
[142] M. Cao, M. Cao, F. Hoffet, S. Qiu, S. Qiu, A. S. Sheremet, A. S. Sheremet, J. Laurat, *Optica, OPTICA* **2020**, *7*, 1440.
[143] K. F. Reim, J. Nunn, V. O. Lorenz, B. J. Sussman, K. C. Lee, N. K. Langford, D. Jaksch, I. A. Walmsley, *Nature Photonics* **2010**, *4*, 218.

[144] M. Hosseini, B. M. Sparkes, G. Campbell, P. K. Lam, B. C. Buchler, *Nature Communications* **2011**, *2*, 174.
[145] I. Novikova, R. L. Walsworth, Y. Xiao, *Laser & Photonics Reviews* **2012**, *6*, 333.
[146] C. Kupchak, T. Mittiga, B. Jordaan, M. Namazi, C. Nölleke, E. Figueroa, *Scientific Reports* **2015**, *5*, 7658.
[147] M. Namazi, C. Kupchak, B. Jordaan, R. Shahrokhshahi, E. Figueroa, *Phys. Rev. Applied* **2017**, *8*, 034023.
[148] M. P. Hedges, J. J. Longdell, Y. Li, M. J. Sellars, *Nature* **2010**, *465*, 1052.
[149] M. Afzelius, C. Simon, *Phys. Rev. A* **2010**, *82*, 022310.
[150] M. Sabooni, Q. Li, S. Kröll, L. Rippe, *Phys. Rev. Lett.* **2013**, *110*, 133604.
[151] F. Bussières, N. Sangouard, M. Afzelius, H. de Riedmatten, C. Simon, W. Tittel, *Journal of Modern Optics* **2013**, *60*, 1519.
[152] M. Gündoğan, P. M. Ledingham, K. Kutluer, M. Mazzera, H. de Riedmatten, *Phys. Rev. Lett.* **2015**, *114*, 230501.
[153] M. Afzelius, C. Simon, H. de Riedmatten, N. Gisin, *Phys. Rev. A* **2009**, *79*, 052329.
[154] H. de Riedmatten, M. Afzelius, M. U. Staudt, C. Simon, N. Gisin, *Nature* **2008**, *456*, 773.
[155] M. Rančić, M. P. Hedges, R. L. Ahlefeldt, M. J. Sellars, *Nature Physics* **2018**, *14*, 50.
[156] S. P. Horvath, M. K. Alqedra, A. Kinos, A. Walther, J. M. Dahlström, S. Kröll, L. Rippe, *arXiv:2006.00943 [quant-ph]* **2020**.
[157] K. T. Kaczmarek, P. M. Ledingham, B. Brecht, S. E. Thomas, G. S. Thekkadath, O. Lazo-Arjona, J. H. D. Munns, E. Poem, A. Feizpour, D. J. Saunders, J. Nunn, I. A. Walmsley, *Phys. Rev. A* **2018**, *97*, 042316.
[158] R. Finkelstein, E. Poem, O. Michel, O. Lahad, O. Firstenberg, *Science Advances* **2018**, *4*, eaap8598.
[159] J.-P. Dou, A.-L. Yang, M.-Y. Du, D. Lao, J. Gao, L.-F. Qiao, H. Li, X.-L. Pang, Z. Feng, H. Tang, X.-M. Jin, *Communications Physics* **2018**, *1*, 1.
[160] J.-S. Tang, Z.-Q. Zhou, Y.-T. Wang, Y.-L. Li, X. Liu, Y.-L. Hua, Y. Zou, S. Wang, D.-Y. He, G. Chen, Y.-N. Sun, Y. Yu, M.-F. Li, G.-W. Zha, H.-Q. Ni, Z.-C. Niu, C.-F. Li, G.-C. Guo, *Nature Communications* **2015**, *6*, 8652.
[161] J. H. Davidson, P. Lefebvre, J. Zhang, D. Oblak, W. Tittel, *Phys. Rev. A* **2020**, *101*, 042333.
[162] N. Akopian, L. Wang, A. Rastelli, O. G. Schmidt, V. Zwiller, *Nature Photonics* **2011**, *5*, 230.
[163] H. Huang, R. Trotta, Y. Huo, T. Lettner, J. S. Wildmann, J. Martín-Sánchez, D. Huber, M. Reindl, J. Zhang, E. Zallo, O. G. Schmidt, A. Rastelli, *ACS Photonics* **2017**, *4*, 868.
[164] J. S. Wildmann, R. Trotta, J. Martín-Sánchez, E. Zallo, M. O'Steen, O. G. Schmidt, A. Rastelli, *Phys. Rev. B* **2015**, *92*, 235306.
[165] T. Kroh, J. Wolters, A. Ahlrichs, A. W. Schell, A. Thoma, S. Reitzenstein, J. S. Wildmann, E. Zallo, R. Trotta, A. Rastelli, O. G. Schmidt, O. Benson, *Scientific Reports* **2019**, *9*, 13728.
[166] H. P. Specht, C. Nölleke, A. Reiserer, M. Uphoff, E. Figueroa, S. Ritter, G. Rempe, *Nature* **2011**, *473*, 190.
[167] A. J. Shields, *Nature Photonics* **2007**, *1*, 215.
[168] J. Wolters, G. Buser, A. Horsley, L. Béguin, A. Jöckel, J.-P. Jahn, R. J. Warburton, P. Treutlein, *Phys. Rev. Lett.* **2017**, *119*, 060502.
[169] M. Fleischhauer, M. D. Lukin, *Phys. Rev. Lett.* **2000**, *84*, 5094.
[170] L. Ma, O. Slattery, X. Tang, *J. Opt.* **2017**, *19*, 043001.
[171] H. M. Meyer, R. Stockill, M. Steiner, C. Le Gall, C. Matthiesen, E. Clarke, A. Ludwig, J. Reichel, M. Atatüre, M. Köhl, *Phys. Rev. Lett.* **2015**, *114*, 123001.
[172] C. H. Bennett, G. Brassard, C. Crépeau, R. Jozsa, A. Peres, W. K. Wootters, *Physical Review Letters* **1993**, *70*, 1895.

[173] J. Nilsson, R. M. Stevenson, K. H. A. Chan, J. Skiba-Szymanska, M. Lucamarini, M. B. Ward, A. J. Bennett, C. L. Salter, I. Farrer, D. A. Ritchie, A. J. Shields, *Nature Photonics* **2013**, *7*, 311.

[174] F. Basso Basset, M. B. Rota, C. Schimpf, D. Tedeschi, K. D. Zeuner, S. F. Covre da Silva, M. Reindl, V. Zwiller, K. D. Jöns, A. Rastelli, R. Trotta, *Phys. Rev. Lett.* **2019**, *123*, 160501.

[175] M. B. Rota, F. B. Basset, D. Tedeschi, R. Trotta, *IEEE Journal of Selected Topics in Quantum Electronics* **2020**, *26*, 1.

[176] M. Zopf, R. Keil, Y. Chen, J. Yang, D. Chen, F. Ding, O. G. Schmidt, *Phys. Rev. Lett.* **2019**, *123*, 160502.

[177] J. Liu, K. Konthasinghe, M. Davanço, J. Lawall, V. Anant, V. Verma, R. Mirin, S. W. Nam, J. D. Song, B. Ma, Z. S. Chen, H. Q. Ni, Z. C. Niu, K. Srinivasan, *Physical Review Applied* **2018**, *9*, DOI 10.1103/PhysRevApplied.9.064019.

[178] C. Schimpf, M. Reindl, F. Basso Basset, K. D. Jöns, R. Trotta, A. Rastelli, *Appl. Phys. Lett.* **2021**, *118*, 100502.

[179] A. V. Kuhlmann, J. H. Prechtel, J. Houel, A. Ludwig, D. Reuter, A. D. Wieck, R. J. Warburton, *Nature Communications* **2015**, *6*, DOI 10.1038/ncomms9204.

[180] F. Olbrich, J. Höschele, M. Müller, J. Kettler, S. Luca Portalupi, M. Paul, M. Jetter, P. Michler, *Appl. Phys. Lett.* **2017**, *111*, 133106.

[181] M. Anderson, T. Müller, J. Skiba-Szymanska, A. B. Krysa, J. Huwer, R. M. Stevenson, J. Heffernan, D. A. Ritchie, A. J. Shields, *Appl. Phys. Lett.* **2021**, *118*, 014003.

[182] T. Müller, J. Skiba-Szymanska, A. B. Krysa, J. Huwer, M. Felle, M. Anderson, R. M. Stevenson, J. Heffernan, D. A. Ritchie, A. J. Shields, *Nature Communications* **2018**, *9*, 862.

[183] E. Saglamyurek, J. Jin, V. B. Verma, M. D. Shaw, F. Marsili, S. W. Nam, D. Oblak, W. Tittel, *Nature Photonics* **2015**, *9*, 83.

[184] R. Riedinger, S. Hong, R. A. Norte, J. A. Slater, J. Shang, A. G. Krause, V. Anant, M. Aspelmeyer, S. Gröblacher, *Nature* **2016**, *530*, 313.

[185] F. Basso Basset, M. Valeri, E. Roccia, V. Muredda, D. Poderini, J. Neuwirth, N. Spagnolo, M. B. Rota, G. Carvacho, F. Sciarrino, R. Trotta, *Sci Adv* **2021**, *7*, DOI 10.1126/sciadv.abe6379.

[186] C. Schimpf, M. Reindl, D. Huber, B. Lehner, S. F. C. D. Silva, S. Manna, M. Vyvlecka, P. Walther, A. Rastelli, *Science Advances* **2021**, *7*, eabe8905.